\documentclass[preprintnumbers,10pt,nofootinbib]{revtex4}

\usepackage{amsmath,latexsym,amssymb,amsfonts}
\usepackage{graphicx,color}
\usepackage{epstopdf}
\usepackage{bm}
 \usepackage{ulem} 

\begin{document}

\preprint{YITP-23-70}

\title{\textbf{Yang-Mills instantons as the end point of black hole evaporation}}

\author{
Pisin Chen$^{a,b,c,d}$\footnote{{\tt pisinchen@phys.ntu.edu.tw}},
Xiao Yan Chew$^{e}$\footnote{{\tt xiao.yan.chew@just.edu.cn}},
Misao Sasaki$^{a,f,g}$\footnote{{\tt misao.sasaki@ipmu.jp}}
and
Dong-han Yeom$^{a,h,i}$\footnote{{\tt innocent.yeom@gmail.com}}
}

\affiliation{
$^{a}$Leung Center for Cosmology and Particle Astrophysics, National Taiwan University, Taipei 10617, Taiwan\\
$^{b}$Department of Physics, National Taiwan University, Taipei 10617, Taiwan\\
$^{c}$Graduate Institute of Astrophysics, National Taiwan University, Taipei 10617, Taiwan\\
$^{d}$Kavli Institute for Particle Astrophysics and Cosmology,
SLAC National Accelerator Laboratory, Stanford University, Stanford, California 94305, USA\\
$^{e}$School of Science, Jiangsu University of Science and Technology, Zhenjiang, 212100 China \\
$^{f}$Kavli Institute for the Physics and Mathematics of the Universe (WPI), University of Tokyo, Chiba 277-8583, Japan\\
$^{g}$Yukawa Institute for Theoretical Physics, Kyoto University, Kyoto 606-8502, Japan\\
$^{h}$Department of Physics Education, Pusan National University, Busan 46241, Republic of Korea\\
$^{i}$Research Center for Dielectric and Advanced Matter Physics, Pusan National University, Busan 46241, Republic of Korea
}

\begin{abstract}
Non-perturbative contributions of the Euclidean path integral are important to understand the information loss paradox. In this paper, we revisit the Yang-Mills instantons in the Einstein-Yang-Mills theory. There exists a globally regular solution that is known as the Bartnik-McKinnon solution and a black hole solution. The regular and the black hole solutions are smoothly connected in the small horizon limit. 
Their Euclidean action is solely characterized by the ADM mass, and the transition probability follows the usual Bekenstein-Hawking entropy formula. Therefore, the Yang-Mills instantons provide a non-perturbative channel to the black hole evaporation, which competes effectively with perturbative processes, and becomes dominant toward the end of evaporation.
We show that these instantons provide a smooth transition mechanism from a black hole to regular spacetime. 
\end{abstract}

\maketitle

\newpage

\tableofcontents

\section{Introduction}

The evaporation of black holes is a consequence of general relativity and quantum mechanics \cite{Hawking:1975vcx}. 
It poses, however, a serious challenge about the consistency between these two fundamental theories since it appears to imply the violation of unitarity.
In other words, a consistent description of an evaporating black hole demands the reconciliation of the tension between general relativity and unitary quantum mechanics \cite{Yeom:2009zp}. Recently, many important and interesting solutions are reported in the literature \cite{Almheiri:2019hni}. The common features of them indicate that the following two conditions are required to explain the unitarity of black hole evaporation \cite{Chen:2021jzx}:
\begin{itemize}
\item[--] 1. \textit{Multi-history condition}: One must take account of the contribution of different semi-classical histories to the wavefunction of the system, which include information-preserving histories, i.e., geometries without event horizons.
\item[--] 2. \textit{Late-time dominance condition}: The contribution of information-preserving histories should dominate the wave function at the late time of the black hole evaporation.
\end{itemize}
If these two conditions are realized, the unitary evolution of the black hole system is automatically guaranteed, and one can consistently derive the Page curve \cite{Page:1993df} for the entanglement entropy \cite{Chen:2021jzx}.

Although the above two conditions look conceptually simple, they are practically highly non-trivial to realize. First, one needs a formulation that is naturally endowed with the multi-history condition which includes information-preserving histories. One possibility is to take account of replica wormholes~\cite{Almheiri:2019hni}. In this case, the two conditions seem realizable by including replica wormhole configurations in the path integral. However, since the computations of replica wormholes are based on the density matrix instead of the wave function, it is unclear whether they can be embedded in the canonical path integral formalism. An alternative is to develop a method to compute the contribution of trivial geometries, i.e., those with neither horizon nor singularity, in the evolution of the system \cite{Hawking:2005kf}. In this case, a crucial issue is to exhibit the late-time dominance condition. In our recent papers \cite{Chen:2021jzx}, we showed that the probability of tunneling toward trivial geometries will dominate as the black hole approaches the end point of evaporation, and hence the late-time condition is realized, under the assumption that instantons that mediate tunneling from the black hole geometry to a trivial geometry exist in general. A price to pay was the modification of the Page curve in such a way that it no longer respects the Bekenstein-Hawking entropy bound~\cite{Chen:2014jwq}.

It is fair to say that, at this moment, the final answer is not yet in sight. It is clear, however, that there is more to be learned about non-perturbative tunneling channels that may contribute to the black hole evaporation. 
In this paper, we take the second, canonical wavefunction approach, and discuss non-perturbative tunneling channels in the Einstein-Yang-Mills (EYM) theory, with the attempt to shed some lights on the important issue of information loss paradox.

An example of a non-perturbative tunneling channel was discussed under the thin-shell approximation, namely, a matter field that allows for a thin spherical shell configuration~\cite{Sasaki:2014spa}. This is a nice example of non-perturbative tunneling channels to trivial geometries, but it is valid only under the assumption of a highly tuned potential, and hence the generality is lost. Another interesting example is to consider a black hole plus free scalar field instantons \cite{Chen:2018aij}, where the Hawking radiation is described as tunneling mediated by these instantons in the small back-reaction limit.\footnote{The idea of describing the Hawking radiation as tunneling seems quite old. See, e.g., \cite{Massar:1997ic}. We thank Ted Jacobson for letting us know about this reference.} However, while we should be able to consistently describe not only the perturbative Hawking radiation but also the non-perturbative tunneling to trivial geometries, explicit construction of these instantons that includes back-reaction remains a technical challenge, as one has to deal with the space-time dependent dynamics of non-perturbative complex-valued instantons.

In this context, it is of great interest if there exists a fully nonlinear instanton solution that can describe the non-perturbative tunneling of a black hole to a trivial geometry. In this paper, with this point of view in mind, we revisit the EYM instantons in the light of the black hole evaporation. Since Yang-Mills (YM) fields surely exist in Nature, 
such instanton solutions are expected to make an important contribution to black hole evaporation. 
 
It has been shown that there are no globally regular solutions in the pure YM field \cite{Deser:1976wq,Coleman:1975qj,Deser:1983fk}. However, Bartnik and McKinnon discovered that regular and asymptotically flat solutions exist once gravity is taken into account, that is, in EYM theory \cite{Bartnik:1988am}. Subsequently, a black hole solution with a non-trivial YM field configuration was also constructed \cite{Bizon:1990sr,Volkov:1990sva,Kuenzle:1990is}. These two solutions are connected smoothly in the small horizon limit. Furthermore, they possess infinite sequences of solutions, which can be labeled as $k$ to represent the higher nodes for the excitation of the gauge field and approach the extremal Reissner-Nordstrom black hole for $k \rightarrow \infty$. It is important to mention that they are unstable against linear or nonlinear perturbations\cite{Straumann:1989tf,Straumann:1990as,Zhou:1991nu,Bizon:1991nt,Bizon:1991hw,Galtsov:1991du,Galtsov:1991nk,Lavrelashvili:1994rp,Volkov:1994dq,Mavromatos:1995kc,Brodbeck:1994vu}. For a more detailed review, see, e.g., Galtsov and Volkov \cite{Volkov:1998cc}. Thus they will eventually decay into particles and radiation.

The fact that they are actually saddle points of the Euclidean action was pointed out by Moss and Wray~\cite{Moss:1992sb}, which implies that they can be legitimately regarded as instantons that contribute to quantum tunneling.
The question is then whether they contribute to the black hole evaporation.
By computing the bounce action, we argue that this is indeed the case. 
In particular, we conclude that the regular Yang-Mills instantons (and their eventual decay products) 
constitute part of the information-preserving histories.
We note that, from the point of view of the Page curve, as these instantons somewhat enhance the probability of trivial geometries relative to the black hole geometry, the violation of the Bekenstein-Hawking bound is slightly eased, though the modification will is significant only toward the end point of evaporation.


This paper is organized as follows. In Sec.~\ref{sec:mod}, we discuss our model and solutions of EYM theory. In Sec.~\ref{sec:app}, we evaluate the Euclidean action of the solutions; based on this computation, we conclude that they indeed play the role of instantons mediating the evaporation process, and they contribute not only near the endpoint of the evaporation but also in a large black hole background. Finally, in Sec.~\ref{sec:imp} and Sec.~\ref{sec:con}, we discuss in light of the information loss paradox and discuss possible future research directions.

\section{\label{sec:mod}Model}

In this section, we review the Einstein-Yang-Mills instantons and discuss their physical properties.

\subsection{Einstein-Yang-Mills theory}

We consider the Einstein-Yang-Mills (EYM) theory with the SU(2) non-Abelian gauge group:
\begin{eqnarray}
S = \int \left[ \frac{R}{16 \pi G } - \frac{1}{2e^2} \text{Tr} \left(F_{\mu\nu} F^{\mu\nu} \right) \right] \sqrt{-g} dx^{4},
\end{eqnarray}
where $R$ is the Ricci scalar, $e$ is the coupling constant of the YM theory, and ${F_{\mu\nu}}$ is the field strength for the non-Abelian gauge field ${A_{\mu}}$ which is defined as
\begin{eqnarray}
{F_{\mu\nu}} &=& \nabla_{\mu}{A_{\nu}} - \nabla_{\nu}{A_{\mu}}
+ i [A_{\mu}, A_{\nu}],\\
A_\mu &=& \frac{1}{2} \tau^a A^a_\mu \,,
\end{eqnarray}
where $\tau_{a}$ is the Pauli matrices. 

Taking variations with respect to the metric $g_{\mu\nu}$ and the non-Abelian gauge field ${A_{\mu}}$
yields the Einstein equation and the Yang-Mills equation, respectively,
\begin{eqnarray}
R_{\mu\nu} -\frac{1}{2} g_{\mu \nu} R &=&  
\frac{\kappa}{e^2} \,\text{Tr} \left( {F_{\mu\alpha}} {F_{\nu}^{\, \alpha}} - \frac{1}{4} g_{\mu\nu}{F_{\alpha\beta}} {F^{\alpha\beta }} \right),
\label{Einsteineq}\\
\nabla_{\mu}F^{\mu\nu} + i \left[ {A_{\mu}}, F^{\mu\nu} \right] &=& 0 \,.
\end{eqnarray}
where $\kappa=8\pi G$. In the following, we use the Planck units, $G=1$, and set $e=1$. Note that the $e \rightarrow 0$ limit corresponds to the decoupling limit of the Yang-Mills field; $e \gg 1$ corresponds to the strong coupling limit and $e \ll 1$ corresponds to the weak coupling limit.

We employ a spherically symmetric ansatz for the metric,
\begin{eqnarray}
ds^{2} &=& - \left( 1 - \frac{2m(r)}{r} \right) \sigma(r) dt^{2} + \frac{dr^{2}}{1 - \dfrac{2m(r)}{r}} + r^{2} \left( d\theta^{2} + \sin^{2} \theta d\varphi^{2} \right)\,, \label{metric_ansatz}
\end{eqnarray}
and a purely magnetic gauge field $(A_t=0)$ with the spherically symmetric property to construct the solutions in the EYM theory,
\begin{eqnarray}
  A_\mu dx^\mu  = \frac{1-w(r)}{2} \left(  \tau_\varphi d\theta - \tau_\theta \sin \theta d\varphi  \right)   \,,   \label{nonAbelian_gauge}
\end{eqnarray}
where $m$, $\sigma$, and $w$ are the functions of $r$. The function $m(r)$ is known as the Misner-Sharp mass \cite{Misner:1964je}, where it provides the ADM mass $M$ at the spatial infinity, i.e., $m(\infty)=M$.

The substitution of Eqs. \eqref{metric_ansatz} and \eqref{nonAbelian_gauge} into the Einstein and Yang-Mills equations yields the following ordinary differential equations (ODEs) for $m$, $\sigma$, and $w$:
\begin{eqnarray}
m' &=& \frac{\kappa}{2} \left( 1 - \frac{2m}{r} \right) w'^{2} + \kappa \frac{\left( w^{2} - 1 \right)^{2}}{4r^{2}}, \label{ode1} \\
\sigma' &=& \frac{2 \kappa \sigma}{r} w'^{2}, \label{ode2} \\
w'' &=& \frac{w'}{2r^{3} \left( 1 - \dfrac{2m}{r} \right)} \left( \kappa \left( w^{2} - 1 \right)^{2} - 4r m \right) + \frac{w \left( w^{2} - 1 \right)}{r^{2} \left( 1 - \dfrac{2m}{r} \right)}.
\label{ode3}
\end{eqnarray}

\subsection{Solutions}

There exist two regular and asymptotically flat solutions in the EYM theory. The first solution is known as the Bartnik-Mckinnon (BM) solution that is entirely regular and does not have a horizon at the center; the second solution is a black hole with a non-trivial Yang-Mills hair outside the event horizon. We need to solve the ODEs with two different sets of boundary conditions in order to construct them numerically.

To construct the BM solution numerically, we integrate the ODEs from the origin $(r=0)$ to infinity. All functions have to be regular at the center $(r=0)$. The requirement that all functions have to be regular at the center
($r = 0$) is attained by requiring the functions to be finite and their
first-order derivatives vanish at $r = 0$; thus a series expansion of
a regular solution at $r=0$ takes the form:
\begin{eqnarray}
m(r) &=& \kappa \overline{W}^2_2 r^3 + \frac{4}{5} \kappa \overline{W}^3_2 r^5 + \mathcal{O}\left(r^7\right)  \,,  \\
\sigma(r) &=& \sigma_0 - 2 \kappa \overline{W}^2_2 r^2  + \mathcal{O}\left(r^4\right)   \,, 
\label{sigmaeq}\\
w(r) &=& 1+ \overline{W}_2 r^2 + \frac{1}{10} \overline{W}^2_2 \left(  3 + 4 \kappa \overline{W}_2  \right) r^4 + \mathcal{O}\left(r^4\right) \,,
\end{eqnarray}
where $\sigma_0$ and $\overline{W}_2$ are the values at the origin. 

For the black hole solution, which we call the Yang-Mills black hole  (YMBH) solution, we integrate the ODEs from the horizon $r_h$ to infinity. Like most BM solutions, all the functions and their derivatives have to be regular at the horizon $r_h$. Imposing these regularity conditions, we also obtain a series expansion of a regular solution at $r_h$ in the form:
\begin{eqnarray}
m(r) &=& m_h+  \frac{\kappa}{4 r^2_h} \left(  1-W^2_h  \right)^2 \left(r-r_h\right) +  \mathcal{O}\left( \left(r-r_h\right)^2 \right)  \,,  \\
\sigma(r) &=& \sigma_h - \kappa \frac{\widetilde{W}^2_1}{r_h}  \left(r-r_h\right) +  \mathcal{O} \left( \left(r-r_h\right)^2 \right)    \,, \\
w(r) &=&  W_h  +  \widetilde{W}_1 \left(r-r_h\right) +  \mathcal{O} \left( \left(r-r_h\right)^2 \right)   \,,
\end{eqnarray}
where $m_h=r_h/2$ is the Misner-Sharp mass at the horizon (if there is no horizon, $r_{h} = m_{h} = 0$), $\sigma_h$ and $W_h$ are the values at the horizon, and $\widetilde{W}_1$ is given by
\begin{equation}
 \widetilde{W}_1 = \frac{2 r_h W_h \left( 1 - W^2_h \right)}{ \kappa \left( 1 - W^2_h   \right)^2 -2 r^2_h  } \,.
\end{equation} 

At infinity, both the YMBH solution and the BM solution share the same asymptotic behavior 
by requiring the metric functions to be asymptotically flat and the gauge field is bounded to unity.
Namely, both solutions have the asymptotic behavior given by
\begin{eqnarray}
m(r) &=&  M -  \frac{\kappa w^2_1}{2 r^3}  \mp  \frac{\kappa w^2_1 \left( 4 w_1 \pm5 M \right)}{4 r^4}  +   \mathcal{O} \left( \frac{1}{r^5} \right) \,,   \label{inf1} \\
 \sigma(r) &=& 1 - \frac{\kappa w^2_1}{2 r^4}   \mp  \frac{6 \kappa w_1 \left(  w_1 \pm 2M  \right)  }{5 r^5} + \mathcal{O} \left( \frac{1}{r^6} \right)  \,, \label{inf2}  \\
w(r) &=& \pm1 + \frac{w_1}{r} \pm  \frac{3 w_1 \left( w_1 \pm 2M  \right)  }{4 r^2} +  \mathcal{O} \left( \frac{1}{r^3} \right)   \,,
\end{eqnarray}
where $w_1$ is a constant determined by each solution.

Note that there are three parameters at the center ($m_{h} = 0$, $\sigma_{0}$, and $\overline{W}_{2}$ for the BM solution; $m_{h}$, $\sigma_{h}$, and $W_{h}$ for the black hole solution).
The choice of $m_h$ determines $M$, while the other two must be tuned so that $\sigma\to 1$ and $w\to\pm1$ at infinity.
Technically, as clear from Eq. (\ref{ode2}),  $\sigma$ has a constant rescaling degree of freedom.
This is fixed by the asymptotic value $\sigma(\infty) = 1$, which consequently fixes the value of $\sigma_0$.
Therefore, the only non-trivial parameter is $\overline{W}_2$ for the BM solution or $W_{h}$ for the YMBH solution, which is to be determined by the condition $w(\infty)=\pm1$. 
An interesting feature of the BM and the YMBH solutions is that there exists a sequence of solutions, where each solution has a different number of zeros of $w(r)$. We call the solution with $k$ nodes the $k$-th node solution, where $k = 1\,,2\,, 3\,,\cdots$.

\begin{figure}
\centering
\mbox{
(a)
 \includegraphics[angle =-90,scale=0.3]{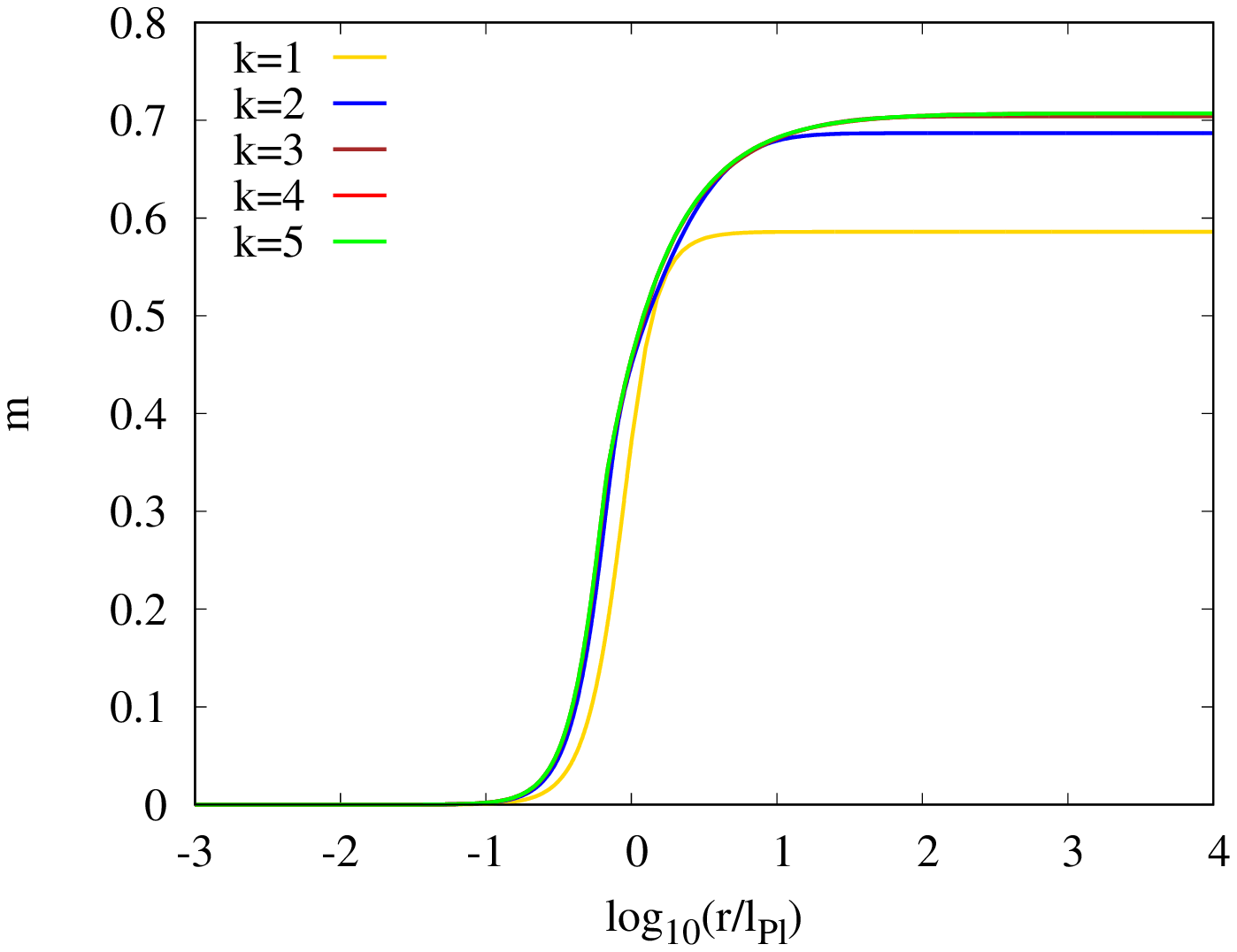}
(b)
 \includegraphics[angle =-90,scale=0.3]{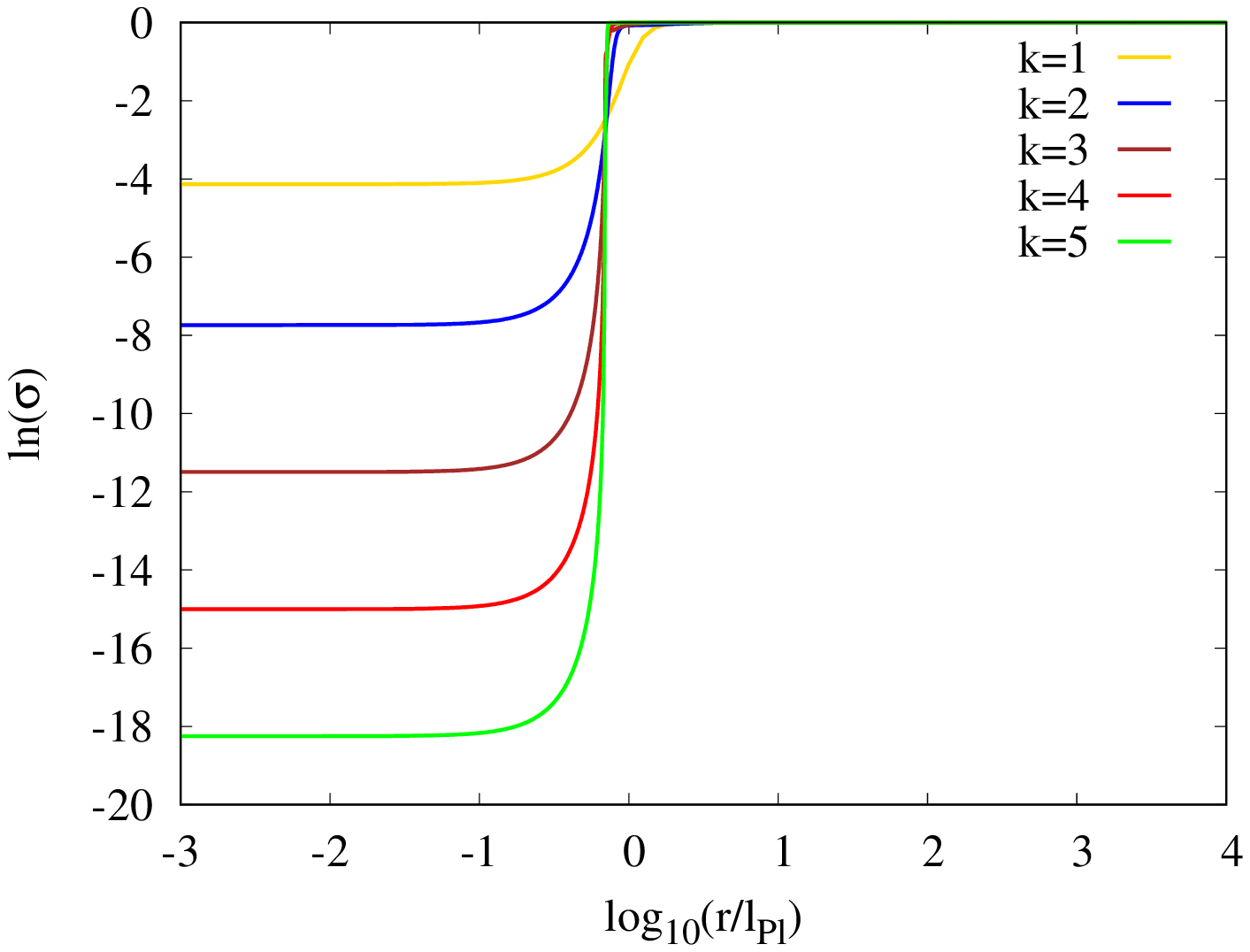}
 }
\mbox{
(c)
 \includegraphics[angle =-90,scale=0.3]{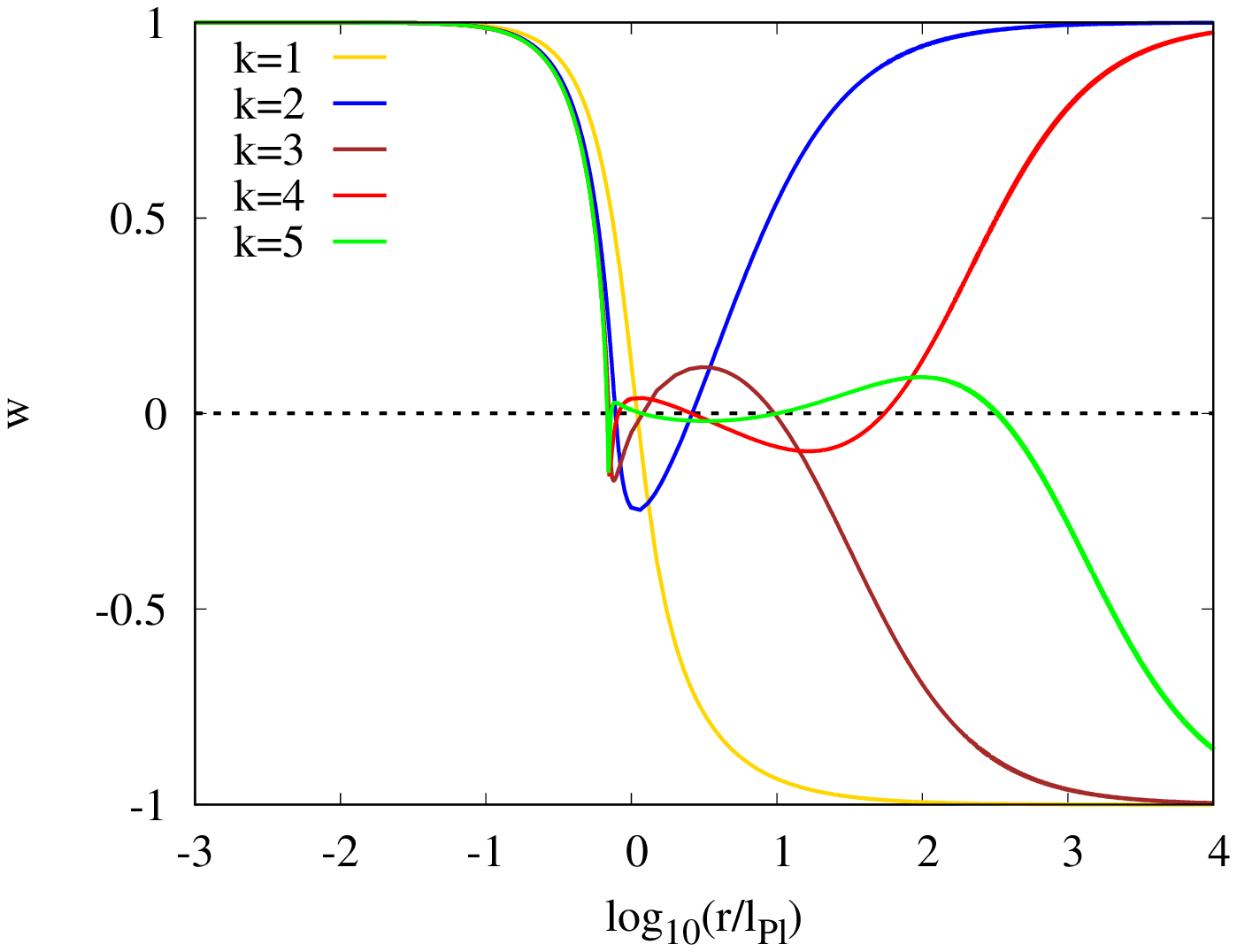}
}
\caption{The profiles of the functions (a) $m(r)$, (b) $\ln(\sigma)$, and (c) $w(r)$, describing the BM solutions with $k$ nodes ($k=1$, 2, 3, 4, 5) in the logarithmic scale of radial coordinate $r$.
}
\label{plot_BM}
\end{figure}

\begin{figure}[h!]
\centering
\mbox{
(a)
 \includegraphics[angle =-90,scale=0.3]{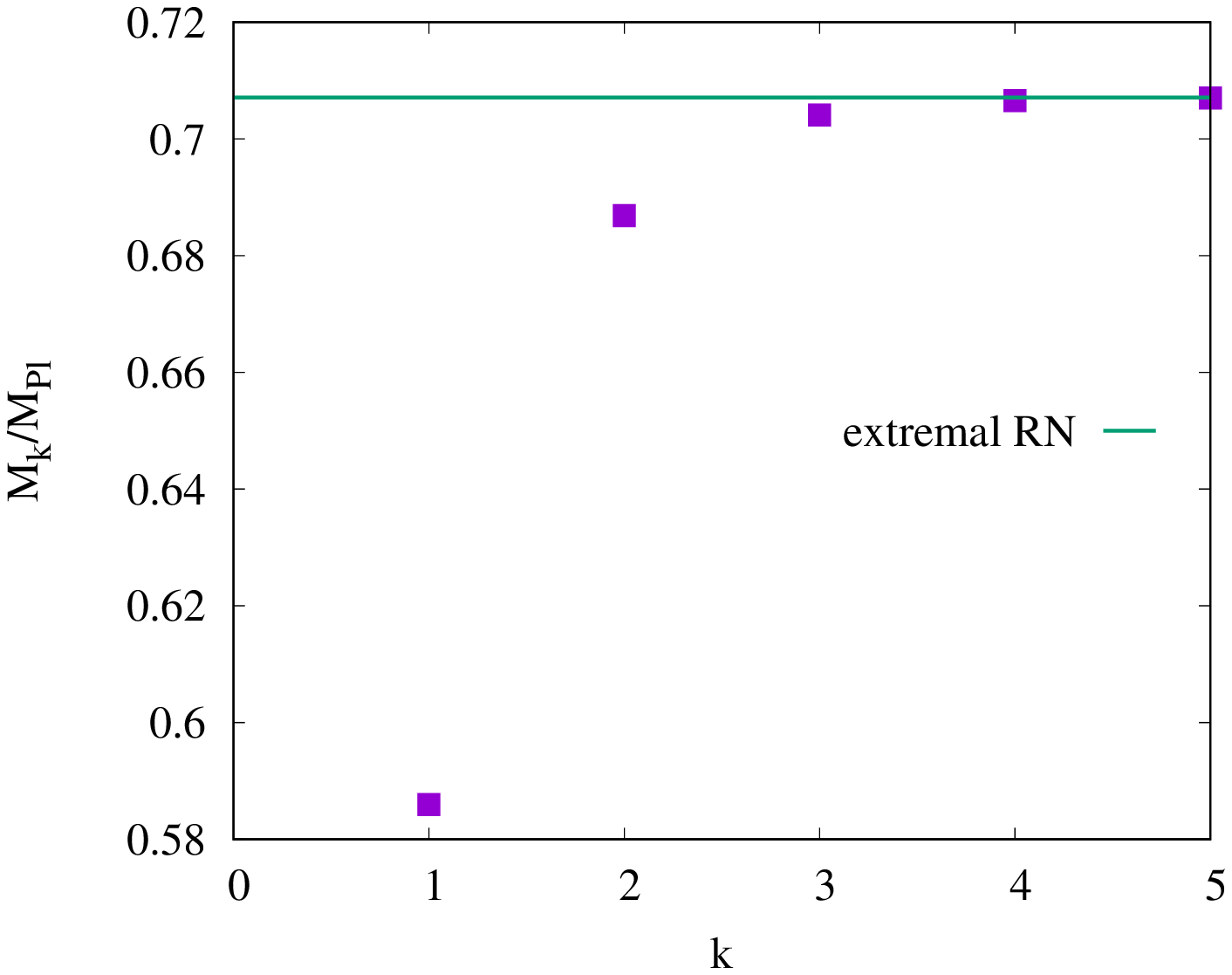}
(b)
 \includegraphics[angle =-90,scale=0.3]{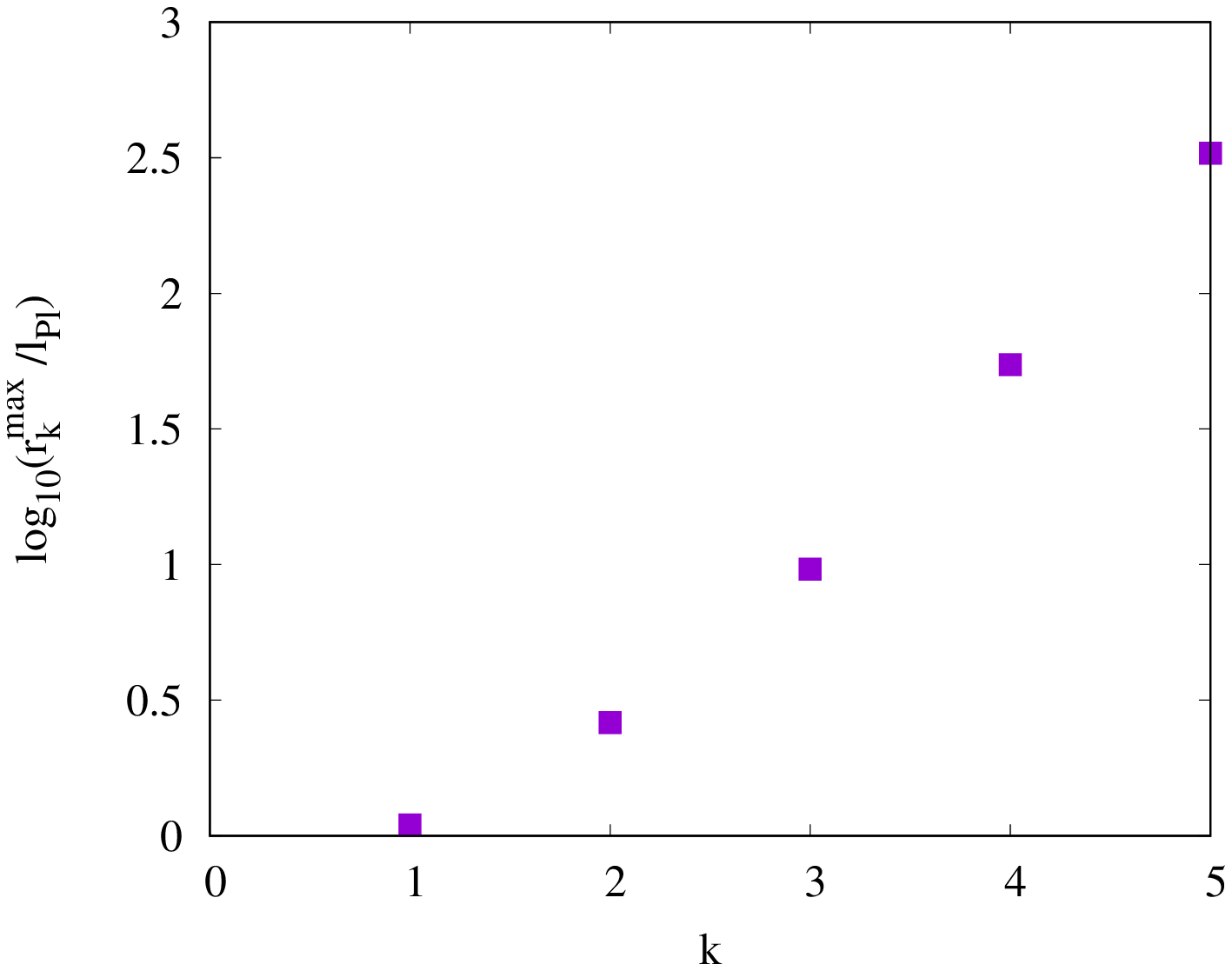}
 }
\caption{Properties of the BM solution: (a) the ADM mass $M_{k}$ and (b) the location $r^{\text{max}}_{k}$ of the outermost zero as a function of $k$.}
\label{plot_BM2}
\end{figure}

\subsubsection{Regular solutions: Bartnik-McKinnon model}

\begin{figure}
\centering
\mbox{
(a)
 \includegraphics[angle =-90,scale=0.3]{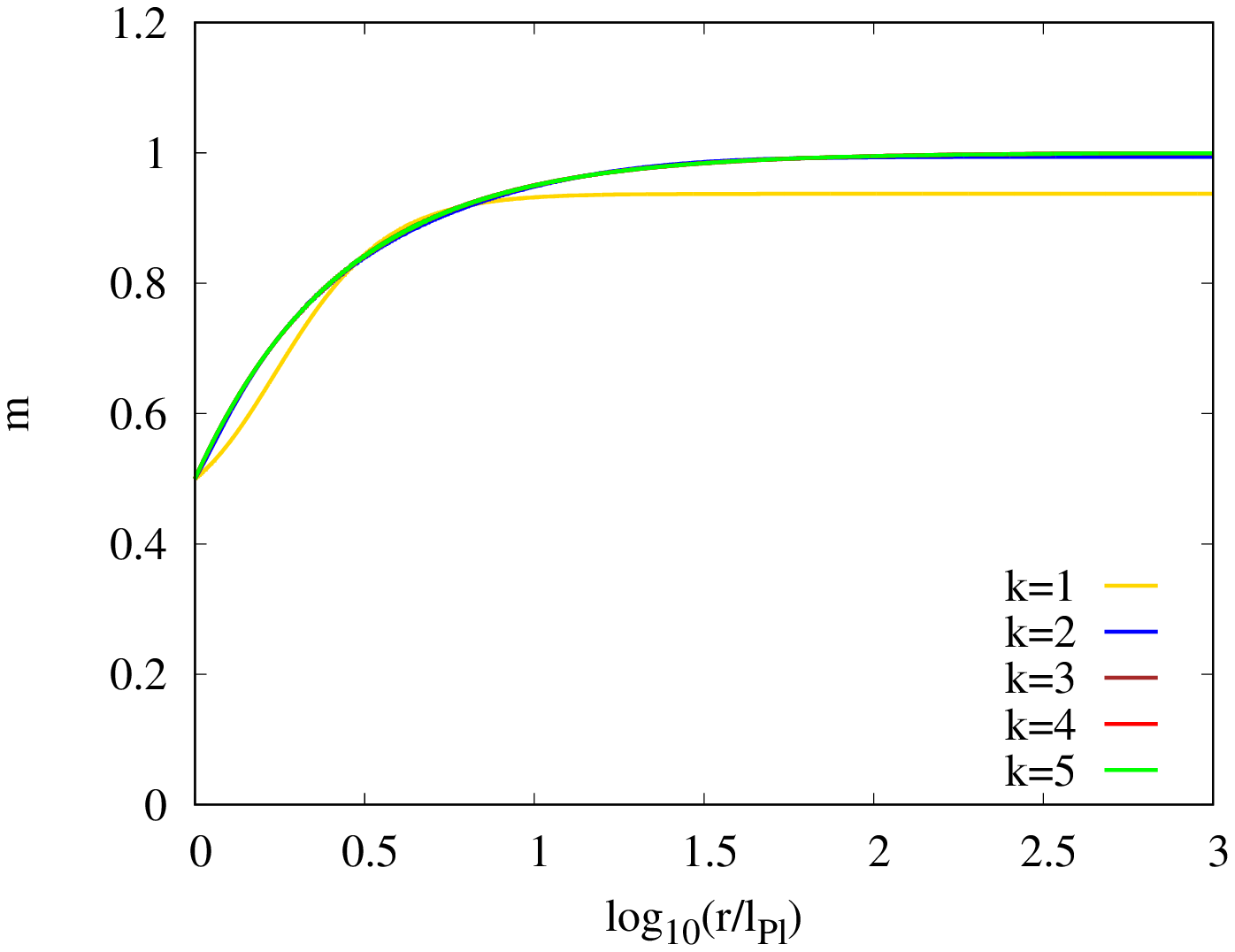}
(b)
 \includegraphics[angle =-90,scale=0.3]{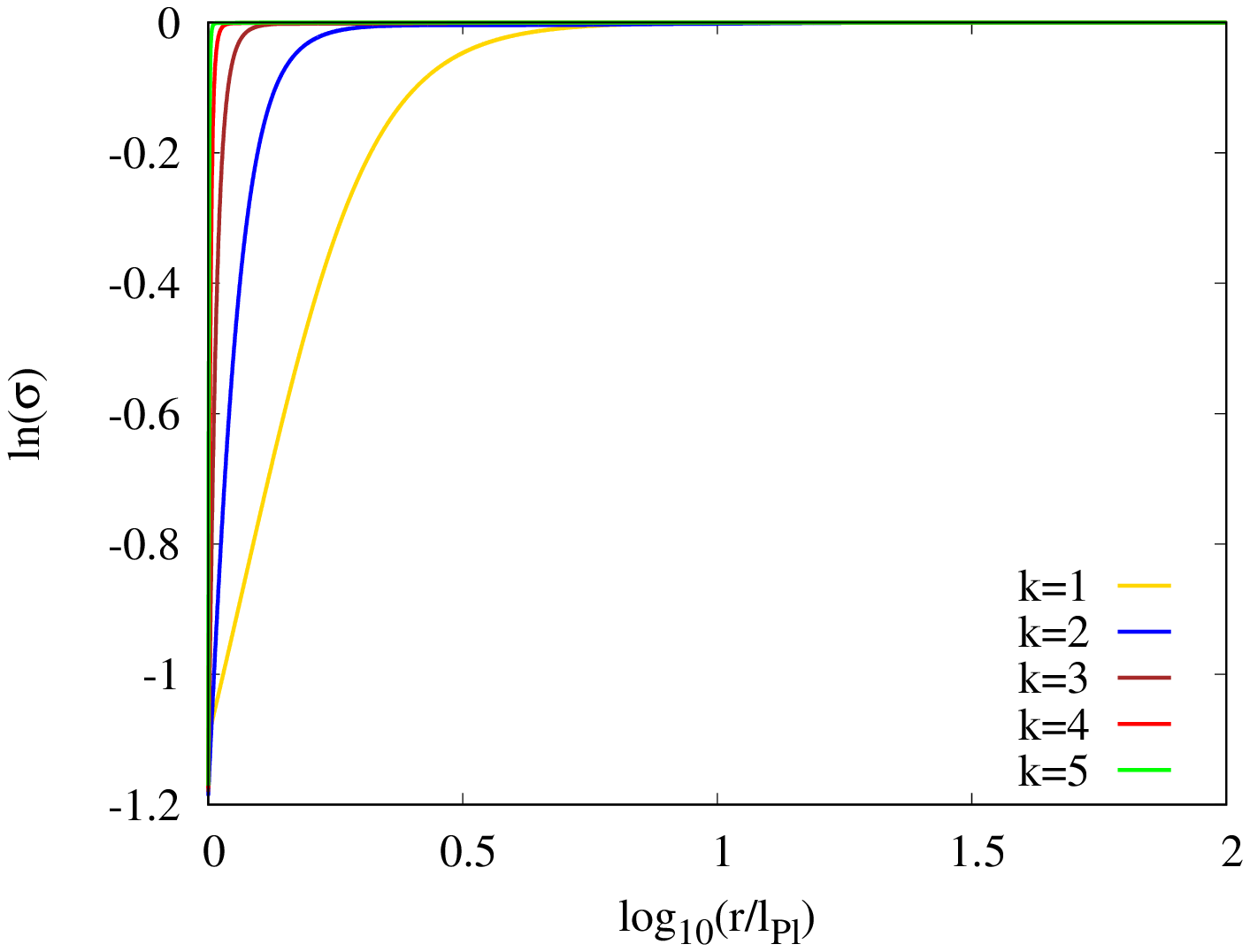}
 }
\mbox{
(c)
 \includegraphics[angle =-90,scale=0.3]{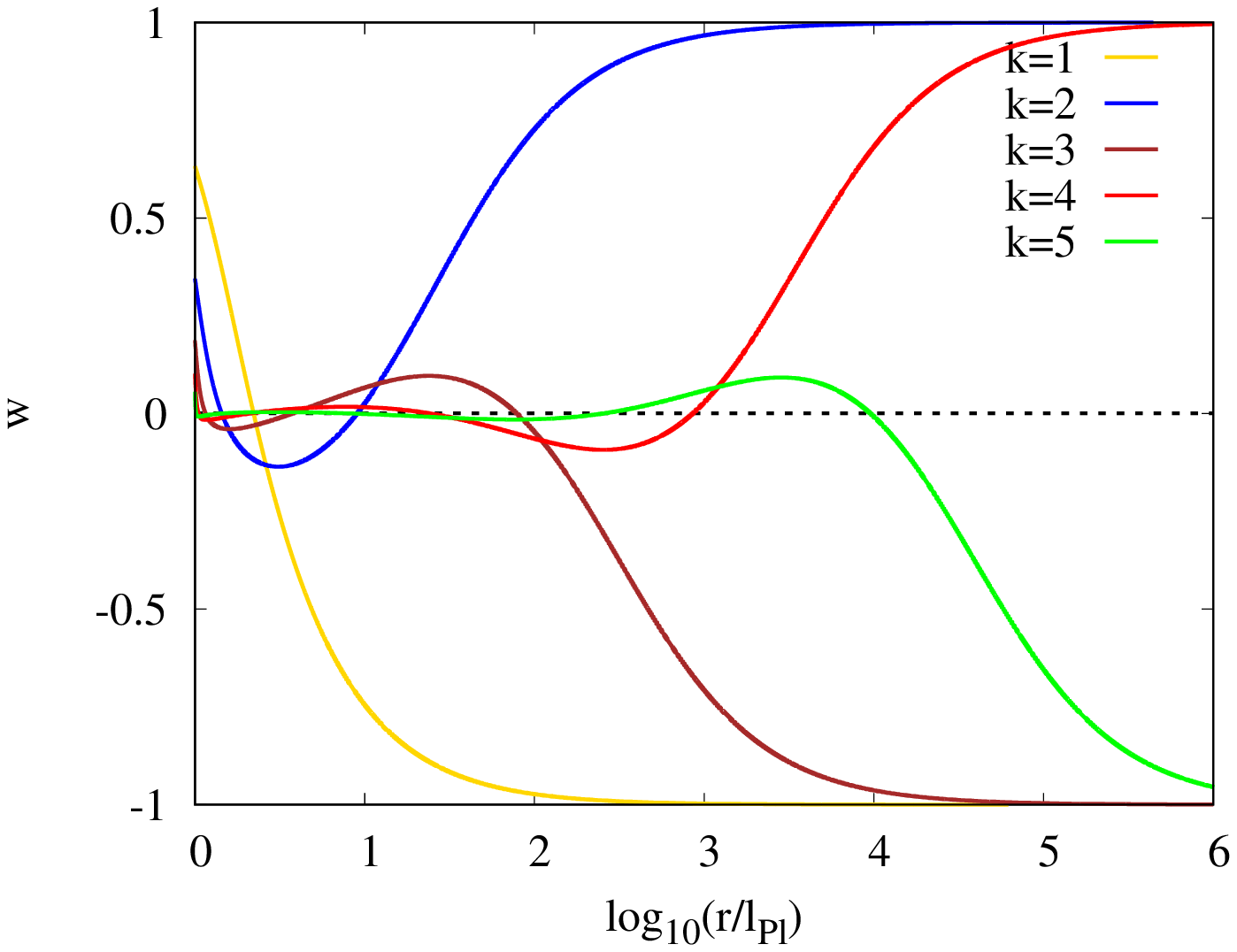}
}
\caption{Profiles of the YMBH solution with $r_h=1$ for $k$ nodes in the logarithmic scale of radial coordinate $r$: (a) $m(r)$, (b) $\ln(\sigma(r))$, (c) $w(r)$.}
\label{plot_YMBH}
\end{figure}

\begin{figure}
\centering
\mbox{
(a)
 \includegraphics[angle =-90,scale=0.3]{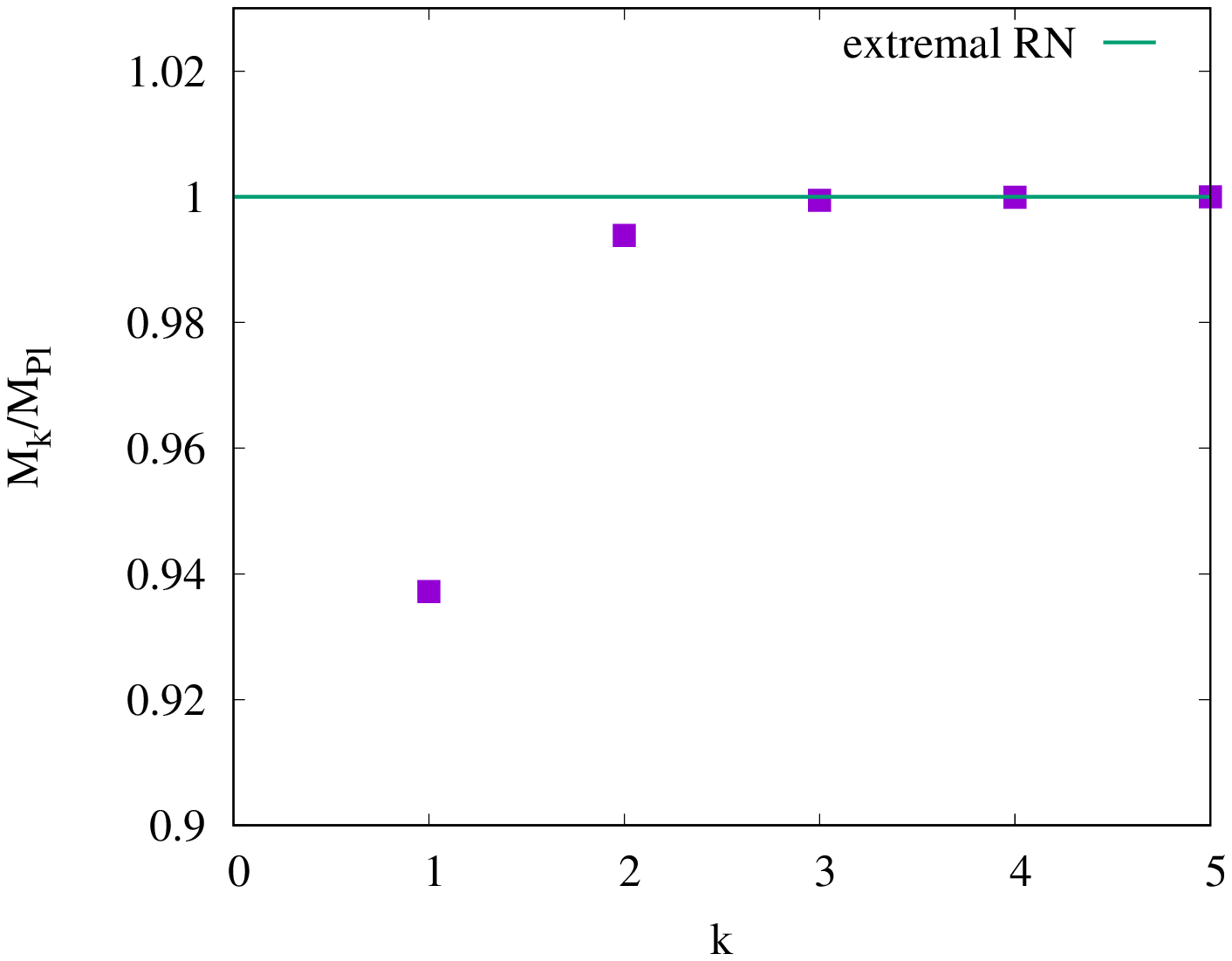}
(b)
 \includegraphics[angle =-90,scale=0.3]{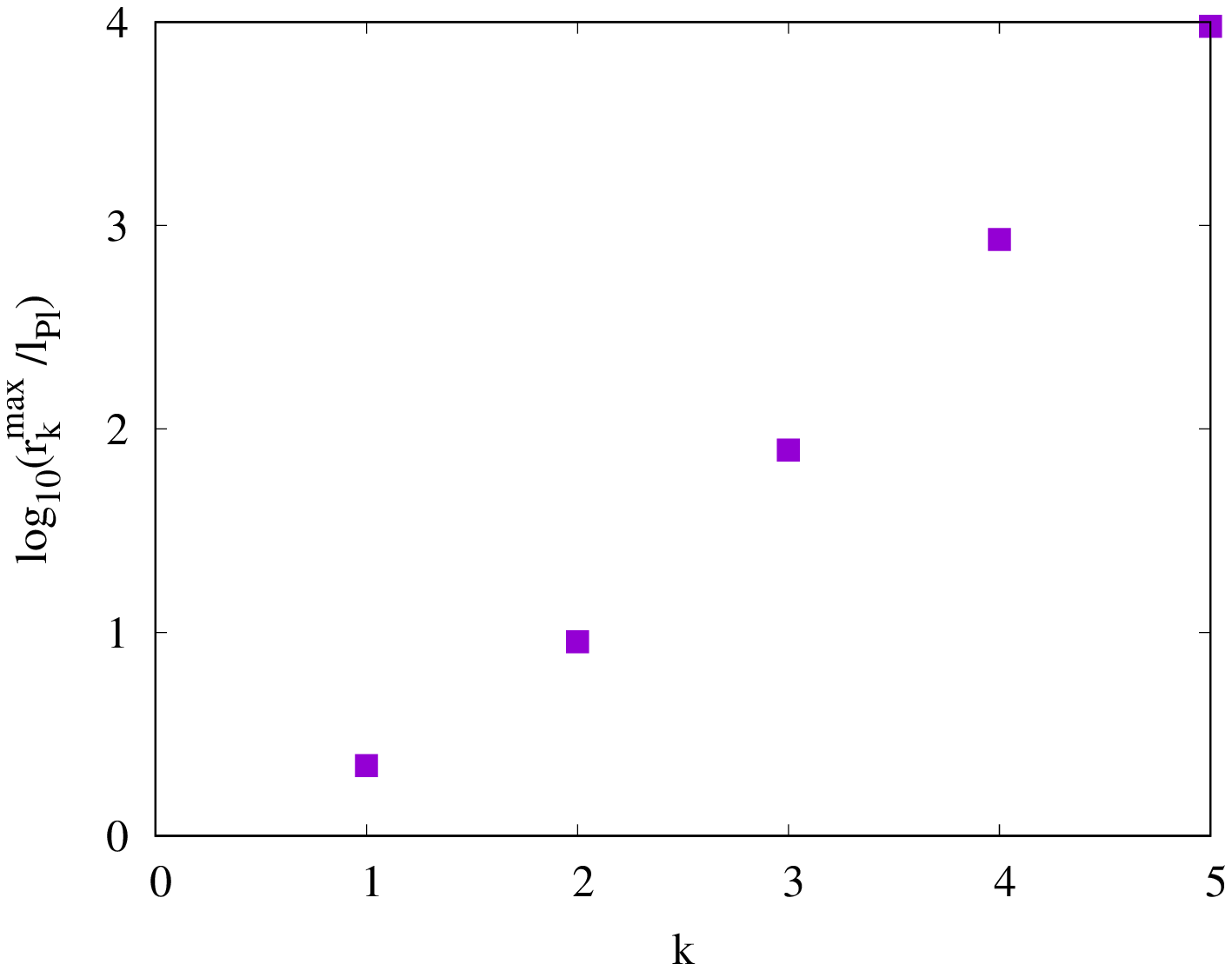}
 }
 \mbox{
 (c)
 \includegraphics[angle =-90,scale=0.3]{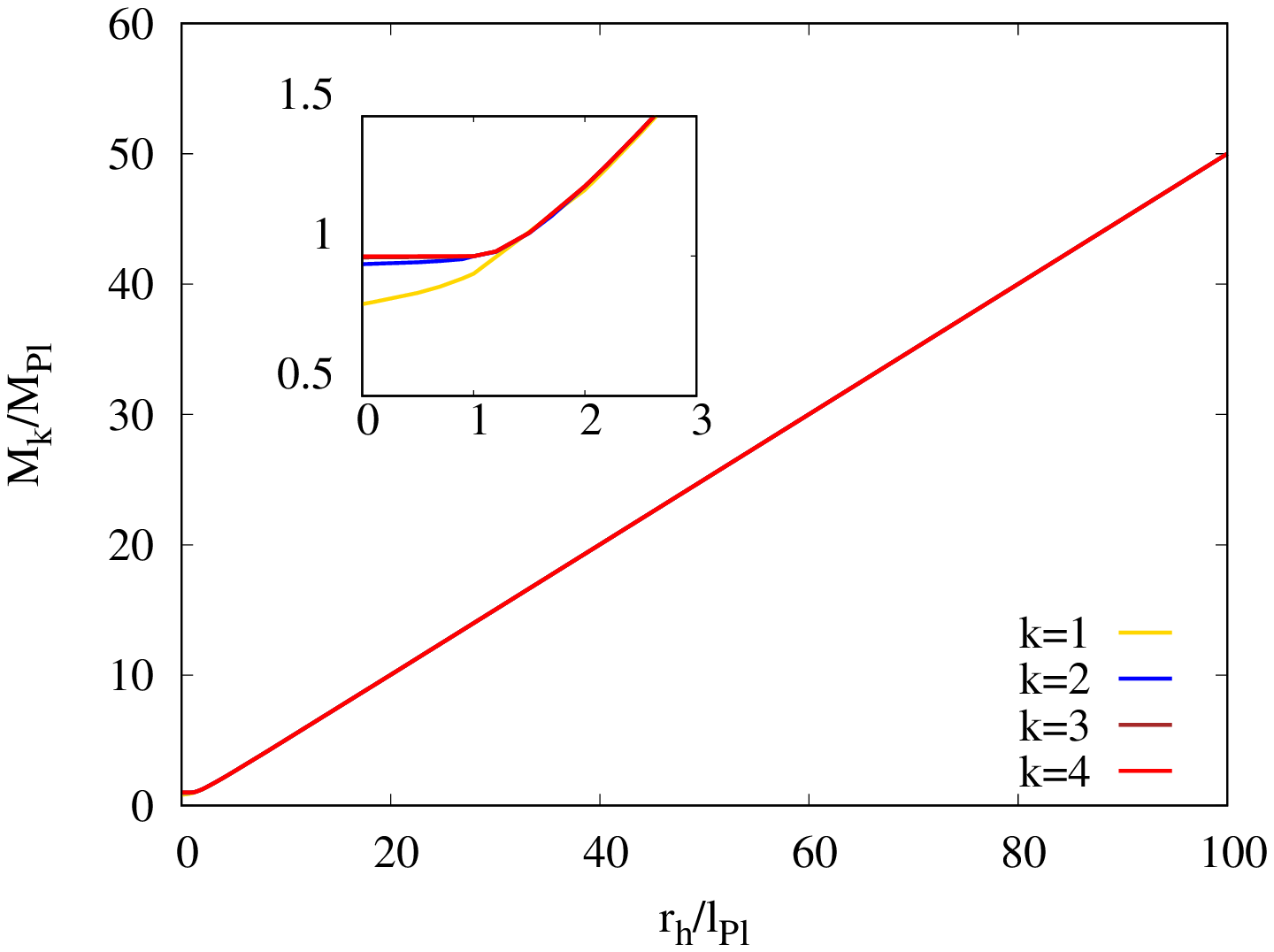}
 (d)
 \includegraphics[angle =-90,scale=0.3]{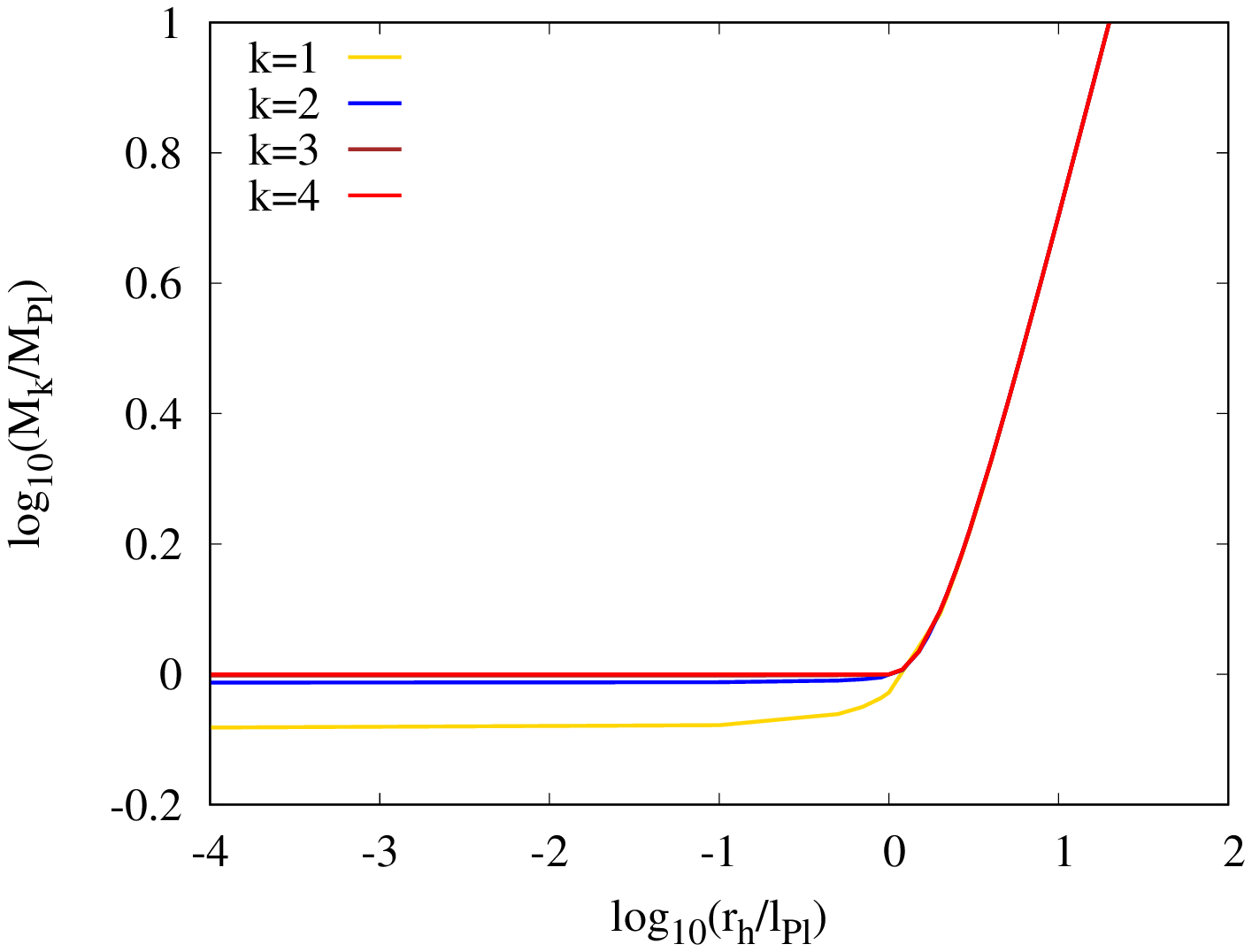}
 }
\caption{Properties of the YMBH solutions with: (a) The mass $M_{k}$ and (b) the locations of the outermost zero, $\text{log}_{10}(r^{\text{max}}_{k})$, as a function of $k$, for $r_h=1$.
(c) The mass $M$ versus the horizon radius $r_h$ for $k=1,\,2,\,3,\,4$, and (d) the same but in a log-log plot.}
\label{plot_YMBH2_plots}
\end{figure}

Let us first review the profiles of the BM solutions for $m(r)$, $w(r)$ and $\sigma(r)$ with nodes $k=1$, 2, 3, 4, 5, in the logarithmic scale of the radial coordinate $r$ in Fig.~\ref{plot_BM}. 
Figs.~\ref{plot_BM}(a) and (b) show that the functions $m(r)$ and $\sigma(r)$ behave monotonically regardless of $k$. Since the BM solution possesses infinite sequences of solutions, then the number of zeros for $w(r)$ increases when $k$ increases as shown in Fig.~\ref{plot_BM}(c). Note that $w(r)$ is bounded in between $\pm 1$. The ADM mass of BM solutions can be read directly from Fig.~\ref{plot_BM2}(a). In the limit of $k\to\infty$, $w\to0$ in the region $1<r<r_k$,  and the geometry there approaches the extremal Reissner-Nordstrom black hole solution, with the asymptotic ADM mass approaching a limiting value of the order of $\sim M_{\mathrm{Pl}}$ as shown in Fig.~\ref{plot_BM2}(a). Fig.~\ref{plot_BM2}(b) shows the location of the outermost zero for each BM solution. If we define the outermost zero of the BM solution as its effective size, then the size of the back hole increases exponentially when $k$ increases linearly, as shown in Fig.~\ref{plot_BM2}(c).

Note that, from the Einstein equations, Eq.~(\ref{Einsteineq}), one readily sees that the effect of $e\neq1$ is to replace $\kappa$ by $\kappa/e^2$.
Then an inspection of Eqs.~(\ref{ode1}), (\ref{ode2}) and (\ref{ode3}) tells us that
the solution can be obtained by the rescaling \cite{Bartnik:1988am},
\begin{eqnarray}
m_{e}(r) = \frac{1}{e} m_{1}(er),
\label{escaling}
\end{eqnarray}
where $m_{e}$ is the solution for an arbitrary $e$ and $m_1$ is the solution for $e = 1$. Since $m_{1}(\infty) \simeq M_{\mathrm{Pl}}$, the ADM mass of the solution increases as $e$ decreases. One can easily notice that, as $e$ goes to zero, the size of the solution increases, and accordingly the mass approaches infinity. Oppositely, if $e$ goes to infinity, the size and the mass of the solution become zero.

\subsubsection{Black holes in Einstein-Yang-Mills Theory}

Similar to the case of the BM solutions, the black holes also possess an infinite sequence of solutions.
Fig.~\ref{plot_YMBH} shows the profiles of the YMBH solutions with nodes $k=1,\, 2,\, 3,\, 4,\, 5$, respectively, for $r_h=1$ in the logarithmic scale of the radial coordinate $r$. 
The similarity also holds for the behaviors of the functions $m(r)$, $\sigma(r)$ and $w(r)$.
Fig.~\ref{plot_YMBH}(a) and (b) show that $m(r)$ and $\sigma(r)$ behave monotonically regardless of $k$, 
 and Fig.~\ref{plot_YMBH}(c) shows that $w(r)$ is bounded in between $\pm 1$ and
the number of zeros for $w(r)$ increases as $k$ increases.
The ADM mass of the black hole solutions can be read off from Fig.~\ref{plot_YMBH}(a). 

In Fig.~\ref{plot_YMBH2_plots}, we show the properties of these infinite sequences of solutions. 
Fig.~\ref{plot_YMBH2_plots}(a) shows the ADM mass as a function of $k$ in the case of $r_h=1$.
Fig.~\ref{plot_YMBH2_plots}(b) shows the locations of the outermost zero of $w$, which increases exponentially with $k$.
Fig.~\ref{plot_YMBH2_plots}(c) exhibits the ADM mass by varying the radius of horizon $r_h$ for 
the solutions with nodes $k=1,\,2,\,3,\,4$.
In the limit $k\to\infty$, the solution also approaches the extremal Reissner-Nordstrom black hole. 
The difference is that we may have an arbitrary large black hole by choosing an arbitrarily large Misner-Sharp mass at the horizon $m_{h}$. 
Hence, the ADM mass approaches the Misner-Sharp mass in the large mass limit, $M\sim m_{h}$ as shown in Figs.~\ref{plot_YMBH2_plots}(c) and (d). 

\begin{figure}
\begin{center}
\includegraphics[scale=0.75]{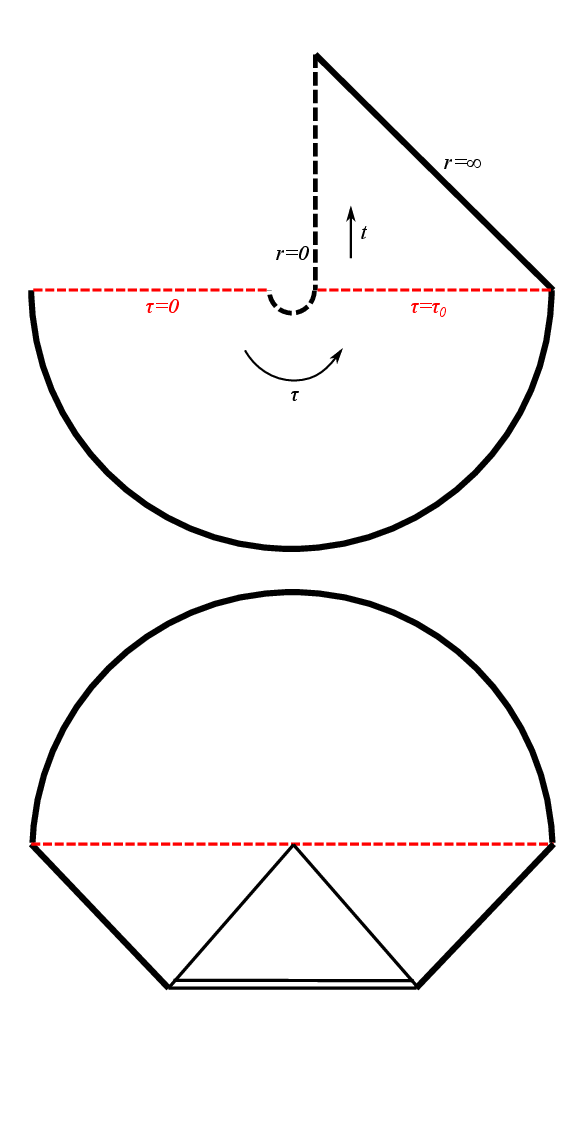}
\caption{\label{fig:int}Schematic diagram of the quantum transition from a black hole to a regular Yang-Mills solution. The red dotted line in the lower figure is the hypersurface through which the Lorentzian black hole is analytically continued to the Euclidean solution. The red dotted lines of the upper figure is the hypersurface through which the Euclidean BM solution is analytically continued to its Lorentzian counterpart. The hypersurfaces $\tau=0$ and $\tau=\tau_0$ are to be identified, with $\tau_0$ being equal to the inverse black hole temperature $\beta$.}
\end{center}
\end{figure}

\section{\label{sec:app}Applications for evaporation}

In this section, regarding the BM and YMBH solutions as instantons, we compute their Euclidean actions. As they are static, the Euclidean solutions are easily obtained by the Wick rotation of the time coordinate. The periodicity in the Euclidean time is arbitrary in the case of the BM solutions, while it is fixed by the Misner-Sharp mass in the case of the black hole solutions.

To describe the process of the initial black hole tunneling to a BM or YMBH solution, we consider the analytic continuations from Lorentzian to Euclidean geometries and vice versa, as illustrated in Fig.~\ref{fig:int}. 
The lower part of Fig.~\ref{fig:int} describes the initial Lorentzian Schwarzschild black hole and its analytic continuation to a Euclidean Schwarzschild spacetime.
It is joined to the instantons in the Euclidean spacetime, and analytically continued back to the Lorentzian geometry as illustrated in the upper part of Fig.~\ref{fig:int}.
We note that since there is neither horizon nor singularity in the BM solution, which is topologically equivalent to that of Minkowski, there is only a single asymptotic infinity.
In order to compute the tunneling probability, following the standard procedure, we evaluate the Euclidean actions of the two Euclidean geometries and subtract each other.

\subsection{Euclidean action}

The Euclidean action is
\begin{eqnarray}
S_{\mathrm{E}} = \int_{\mathcal{M}} \left[ - \frac{R}{16 \pi} + \frac{1}{2} \text{Tr} \left( {F_{\mu\nu}} {F^{\mu\nu}} \right) \right] \sqrt{+g} dx^{4} - \frac{1}{8\pi} \int_{\partial \mathcal{M}} \left( K - K_{0} \right) \sqrt{+h} d^{3}x,
\end{eqnarray}
where $K$ and $K_{0}$ in the second term are the Gibbons-Hawking boundary term of the solution and the Minkowski background, respectively.

The Ricci scalar term is, after the integration by parts,
\begin{eqnarray}
\int \left[ - \frac{R}{16 \pi} \right] \sqrt{+g} dx^{4} &=& 4\pi \beta \int_{r_{h}}^{\infty} dr \sqrt{\sigma} \left( - \frac{2}{\kappa} m' \right) + \frac{\beta}{4} \left.\left( \left( 2m - 2 r m' \right) \sqrt{\sigma} + \left( r-2m \right) r \frac{\sigma'}{\sqrt{\sigma}} \right)\right|_{r_{h}}^{\infty}
\nonumber\\
&=& 4\pi \beta \int_{r_{h}}^{\infty} dr \sqrt{\sigma} \left( - \frac{2}{\kappa} m' \right) + \frac{\beta}{2} \left( M - m_{h} \right),
\label{Ricci-term}
\end{eqnarray}
where $\beta$ is the Euclidean time period. Recall that $r_{h} = m_{h} = 0$ in the absence of a black hole. 
Note that the asymptotic behavior of the solution, Eqs.~\eqref{inf1} and \eqref{inf2}, requires $\sigma\to1$, $\sigma'=O(r^{-5})$ and $m'=O(r^{-4})$ at infinity.

The Yang-Mills term is
\begin{eqnarray}
\int \left[ \frac{1}{2} \text{Tr} \left( {F_{\mu\nu}} {F^{\mu\nu }} \right) \right] \sqrt{+g} dx^{4} &=& 4\pi \beta \int_{r_{h}}^{\infty} dr \sqrt{\sigma} \left( \frac{(1-w^{2})^{2}}{2r^{2}} + \left( 1 - \frac{2m}{r} \right) w'^{2} \right)
\nonumber\\
&=& 4\pi \beta \int_{r_{h}}^{\infty} dr \sqrt{\sigma} \left( \frac{2}{\kappa} m' \right),
\label{YM-term}
\end{eqnarray}
where we have used Eq.~\eqref{ode2} for $m'$ to obtain the second line. 

The contribution from the Gibbons-Hawking boundary term, which is the same as that in the Schwarzschild black hole case, is given by
\begin{eqnarray}
- \frac{1}{8\pi} \int_{\partial \mathcal{M}} \left( K - K_{0} \right) \sqrt{+h} d^{3}x = \beta \frac{M}{2}.
\label{GH-term}
\end{eqnarray}

Adding Eqs.~(\ref{YM-term}) and (\ref{GH-term}), we obtain
\begin{eqnarray}
S_{\mathrm{E}} = \beta M - \frac{\beta}{2} m_{h}\,.
\end{eqnarray}
We note that regarding $\beta$ as the inverse of the temperature determined by the central black hole, $\beta=1/T=8\pi m_h$, this has a natural thermodynamic interpretation.
Namely, since the second term may be expressed as $S=4\pi m_h^2$, which is the entropy of the black hole at the center, we have $TS_{\mathrm{E}}=M-TS$, which is the free energy of the black hole.

\subsection{Interpretation}

\subsubsection{Large black holes}

If one considers the Hawking radiation from a Schwarzschild black hole as tunneling mediated by the above instanton, we should take $\beta=8\pi M$, where $M$ is the total energy of the system that remains the same before and after the tunneling. 
In this case, since the horizon mass is $m_{h} = M - \omega$, where $\omega$ is the energy carried out as Hawking radiation, there appears a cusp at the horizon. However, by using the regularization method \cite{Gregory:2013hja}, one can compute the correction due to the cusp, and 
the difference between the final and initial Euclidean actions is found to be given exactly by the
difference between the initial and final horizon surface areas \cite{Moss:1992sb},
\begin{eqnarray}
B\equiv S_E(M-\omega)-S_E(M)= \frac{\Delta \mathcal{A}}{4} = 4\pi \left( M^{2} - \left( M - \omega \right)^{2} \right) \simeq 8\pi M \omega = \beta \omega,
\end{eqnarray}
where we assumed $\omega \ll M$.

Here a comment is in order. Since there exist infinitely many solutions for a given $M$ with increasing size $r_k$ 
($k=1,2,3.\cdots$), one should sum over all the contributions. 
Let us define $\omega_{k} = M - m_{h,k}$. As $\omega_k$ approaches a constant in the limit $k\to\infty$, 
as discussed in Section \ref{sec:mod}, one might worry that these infinitely many solutions would lead to a divergent contribution to the probability, whereas we have only considered the leading order WKB action so far. The contribution
from the next WKB order is commonly interpreted as the prefactor $A$ of the tunneling rate, that is,
\begin{eqnarray}
\Gamma \simeq A e^{-B}.
\end{eqnarray}
It is known in general that $A$ has the volume dependence \cite{Dunne:2005rt}, 
which is proportional to the energy divided by the volume (see also \cite{Coleman:1978ae}, where $A$ is inversely proportional to the fourth power of the instanton size). 
Therefore, although we do not explicitly compute the prefactor here,
as it is beyond the scope of the present paper,
it is reasonable to expect that the prefactor $A_k$ for the $k$-node solution will provide a large additional 
suppression factor $\sim V_k^{-1}=r_k^{-3}$, since the volume is exponentially large, $V_k\propto e^{C\,k}$, where $C=O(10)$ (see Fig.~\ref{plot_BM2}(b)).


To summarize, the dominant contribution comes only from small $k$'s, and hence we obtain
\begin{eqnarray}
\Gamma_{\mathrm{bh}\rightarrow\mathrm{sol}} \simeq  e^{- \beta \omega}
\end{eqnarray}
as we expected.

\subsubsection{Final stage of black hole evaporation}
The contribution of the regular BM solutions to the evaporation of a black hole of mass $M$ should be important only when $M\simeq M_k$, based on mass conservation.
Since $M_{k}=\mathcal{O}(M_{\mathrm{Pl}})$ for any $k$, they play an important role only near the end of the evaporation, that is, only for Planck-mass scale black holes. Note that since the mass of the BM solution scales as $1/e$, where $e$ is the gauge coupling constant as given in Eq.~(\ref{escaling}), it could be much heavier than the Planck mass if $e\ll1$. However, we expect $e\sim0.1$ as a typical value in the standard model. Thus, it is expected that the mass is somewhat larger than the Planck mass, but not extremely larger.

The tunneling rate due to the $k$-node solution is then approximately
\begin{eqnarray}
\Gamma_k \simeq e^{-\beta M_{k} + 4\pi M^{2}} = e^{-4 \pi M_k^2},
\end{eqnarray}
when $M=M_k$. If all $k$'s would contribute equally, this would mean a catastrophic end of evaporation when the black hole mass becomes equal to that of the BM solution with asymptotically large $k$. However, repeating the same argument as the one in the previous subsection, we do not expect this to happen.
Still, we expect that the regular BM solutions may have a mild influence on the evaporation at the final stage, by contributing a few additional evaporating channels. 
In any case, as the black hole mass approaches the Planck scale, the semi-classical approximation would break down. It is therefore premature to draw any definitive conclusion. 



The final question is, what happens after the tunneling?
The instability of the Yang-Mills instantons is known in the literature \cite{Straumann:1989tf}. On the other hand, it is important to note that the Bartnik-McKinnon solution is a natural generalization of the sphalerons \cite{Galtsov:1991du}. It is therefore reasonable to expect that after they are formed, they would decay into particles. The decay of these instantons may contribute to the baryon number violation.

\section{\label{sec:imp}Implications to the information loss problem}

In order to explain the unitary of the Page curve, let us recall the following two ansatz \cite{Chen:2021jzx}:
\begin{itemize}
\item[--] 1. \textit{Multi-history condition}: For the entire path integral, there must be contributions from various semi-classical histories, some of which are information-preserving.
\item[--] 2. \textit{Late-time dominance condition}: The contribution from the information-preserving histories should dominate at the late time of the black hole evaporation.
\end{itemize}

For simplicity, let us assume that there are only two semi-classical histories; one is the information-losing history, in which the black hole evaporates and vanishes in the end,
and the information-preserving history, in which the black hole tunnels to a trivial geometry mediated by an instanton. 
Let $p_{1}(t)$ be the probability of the information-losing history, which is a function of time, and $p_{2}(t)$ be the probability of the information-preserving history, where $p_{1} + p_{2} = 1$ at each instant of time.

Now consider the entanglement entropy by separating the system into two: the regions inside and outside the black hole. It is natural to expect that $S_{1}$ monotonically increases. Namely, if $S_0 = 4\pi M_0^{2}$ is the initial Bekenstein-Hawking entropy of the black hole and $S=4\pi M^2$ is the Bekenstein-Hawking entropy at some later time, we may assume
\begin{eqnarray}
S_{1}(S) = S_{0} - S\,,
\end{eqnarray}
as long as the Hawking radiation produces maximally entangled pairs \cite{Hwang:2017yxp}. On the other hand, for the information-preserving history, in which there is neither horizon nor singularity, we have
\begin{eqnarray}
S_{2}(S) = 0.
\end{eqnarray}
Then the expectation value of the entanglement entropy of the entire system is given by
\begin{eqnarray}
S=p_{1} S_{1} + p_{2} S_{2}=p_1 (S_0-S)\,.
\end{eqnarray}
It is clear that the entanglement entropy increases at the early stage of the evaporation when $p_1(t)\approx1$, while it decreases to zero toward the end of evaporation as $p_{1}(t)=1-p_2(t) \rightarrow 0$ by assumption. This explains the unitary Page curve.

Then the next question is: what happens if we include additional non-perturbative contributions like those discussed in this paper? 
Note that $p_{2}$ will dominate near the Planck scale. If multiple instantons contribute to the tunneling toward trivial geometries, it is quite reasonable to expect that the probability for $p_{2}$ would be enhanced. This would then render the dominance of information-preserving histories to shift toward an earlier time. Namely, for a given black hole mass, the probability of tunneling to a trivial geometry should be higher with multiple instantons than without. Consequently, the Page time, at which the expectation value of the entanglement entropy starts to decrease, may appear earlier than the case without such non-perturbative channels.




\section{\label{sec:con}Conclusion}

In this paper, we presented an alternative interpretation of the known static Einstein-Yang-Mills solutions as instantons that may contribute to the black hole evaporation. There are two classes of solutions: One is the globally regular Bartnik-Mckinnon (BM) solutions, and the other is the Yang-Mills black holes (YMBHs) that have black holes at the center.
The YMBH solutions smoothly reduce to the BM solutions in the limit of vanishing horizon radius.
It is known that there exists an infinite number of solutions, characterized by the number of nodes in the gauge field. However, we gave a reasonable argument that only the solutions with zero node would make a dominant contribution to the evaporation.

Both BM and YMBH solutions have no gauge charges. They are characterized solely by their ADM mass at spatial infinity. Hence their Euclidean actions agree with the usual entropy formula without conserved charges. This allows us to regard these solutions as instantons that provide a non-perturbative channel to the black hole evaporation, that can compete with perturbative processes at late time. Furthermore, when the mass of an evaporating black hole becomes equal to that of the BM solution, it will provide a smooth transition mechanism from the black hole geometry to a regular spacetime.


To conclude, let us emphasize again the importance of non-perturbative instantons. To demonstrate our point, we focused on the $SU(2)$ gauge theory
in this paper, knowing that there should exist more diverse solutions from larger fundamental gauge symmetries. Even though we still retain the conclusion that the Bekenstein-Hawking entropy bound should be violated in our description, the violation is tamed when non-perturbative solutions are considered as the black hole mass approaches the Planck scale. We will continue to investigate such non-perturbative instantons. 
 
Finally, let us mention that there may be instantons that violate global symmetries, such as the baryon number. The implications of the global charge violation, not only in terms of Euclidean quantum gravity but also in terms of particle physics and cosmology, must be an interesting future research topic.

\newpage

\section*{Acknowledgment}

This work was supported in part by JSPS KAKENHI Grant No. 20H05853.
PC is supported by Taiwan's National Science and Technology Council (NSTC) under project number 110-2112-M-002-031, and by the Leung Center for Cosmology and Particle Astrophysics (LeCosPA), National Taiwan University. DY was supported by the National Research Foundation of Korea (Grant No. : 2021R1C1C1008622, 2021R1A4A5031460). XYC is supported by the starting grant of Jiangsu University of Science and Technology (JUST).


\end{document}